\documentclass[a4paper]{article}
\pdfoutput=1
\usepackage{epic,eepic,amsmath,amssymb,color,graphicx}
\usepackage{rotating}

\newcommand{\red}{\color{red}}
\newcommand{\pmm}{p_{\text{m}}}
\newcommand{\pii}{p_{\text{i}}}
\newcommand{\nc}{N_{\text{c}}}
\newcommand{\nt}{N_{\text{t}}}

\makeatletter
\renewcommand{\@makecaption}[2]{
   \vskip\abovecaptionskip
   \sbox\@tempboxa{#1. #2}%
   \ifdim \wd\@tempboxa >\hsize
      \hspace*{1cm}\begin{minipage}{14cm}{\bf #1.} #2\end{minipage}%\par
   \else
     \global \@minipagefalse
     \hb@xt@\hsize{\hfil\box\@tempboxa\hfil}%
   \fi
   \vskip\belowcaptionskip}
\makeatother

\begin{document}

\begin{flushright}  LAPTH - 1217/07\\
%  Roma ...
\end{flushright}

\vspace*{1cm}
\begin{center}{\huge \textbf{An evolutionary model with Turing machines}}

\vspace{10mm}
\begin{minipage}[]{110mm}\begin{center}
{\large Giovanni Feverati$\mbox{}^{\S}$ \& Fabio Musso$\mbox{}^{\dag}$}\\[10mm]
$\mbox{}^{\S}$ Laboratoire de physique theorique LAPTH,\\
CNRS, UMR 5108, associ\'e \`a l'Universit\'e de Savoie,\\ 
9, Chemin de Bellevue, BP 110, 74941, Annecy le Vieux Cedex, France
feverati@lapp.in2p3.fr\\[2mm]
$\mbox{}^{\dag}$ Dipartimento di fisica, Universit\`a degli studi di Roma TRE\\
and Istituto Nazionale di Fisica Nucleare, sezione di Roma TRE,\\
Via della Vasca Navale 84, 00146 Roma, Italy\\
musso@fis.uniroma3.it
\end{center}
\end{minipage}
\end{center}

\vspace{10mm}
\begin{quotation}
\textbf{Abstract.} 
The development of a large non-coding fraction in eukaryotic DNA and the phenomenon of
the code-bloat in the field of evolutionary computations show a striking similarity.
This seems to suggest that (in the presence of mechanisms of code growth) the evolution of a complex code
can't be attained without maintaining a large inactive fraction. 
To test this hypothesis we performed computer simulations of an evolutionary toy model for Turing machines,
studying the relations among fitness and coding/non-coding ratio while varying mutation and code growth rates.
The results suggest that, in our model, having a large reservoir of non-coding states constitutes a great
(long term) evolutionary advantage.
\end{quotation}

\section{Introduction}

The ``C-value enigma'' refers to the fact that DNA size variation among eukaryote species show no relation to 
number of coding genes or to organismal complexity. The discovery that, in eukaryotes, the great majority of DNA is 
(protein) non-coding, led to a possible solution of the paradox. Indeed if there exists ``non-informative'' DNA, then there is no need of correlation between organismal complexity and the total amount of DNA. 
On the other hand, there is no consensus on the actual fraction of the non-coding part that has to be considered
``non-informative''.  
Moreover, the reason why eukaryotes should accumulate and maintain such a large amount of non-coding DNA is still debated. 
At least four different theories have been formulated to answer to this question: ``junk DNA'', ``selfish DNA'',
``nucleoskelethal'' and ``nucleotypical'' theory (for a review, see 
\cite{Gregory} and the references therein).
According to the first two theories, in eukaryotes there is a upward mutation pressure acting to increase DNA 
content without any direct benefit for the ``host'' (but, on the contrary, with a slight harm). This increasing 
tendency will continue until
the mutation pressure is balanced by natural selection acting on the phenotypic level. 
The next two theories state, viceversa, that DNA size is directly related to cell size 
(in a ``coevolutionary'' or ``causative'' way, respectively)
and it is consequently adjusted by natural selection in such a way to obtain the optimal dimension of the cells.

A common feature of all these theories is that the actual content of the added extra DNA is insignificant. 
Therefore such DNA can be freely mutated and it can be viewed as a reservoir of raw material for the 
production of new genes. The relevance of this process for evolution has been emphasized by Ohno \cite{Ohno} in  
the ``junk DNA'' hypothesis context and by \cite{Jain} in the ``selfish DNA'' one (however for an opposite point of view
see, for example, \cite{Bestor}).

Evolutionary algorithms are stochastic search methods that mimic the language of natural biological 
evolution\cite{fogel}. 
They operate on a population of potential solutions to a given problem applying the principle of survival of 
the fittest to produce 
better and better approximations to a solution. At each generation, a new set of approximations is created by the process of 
selecting individuals according to their level of fitness in the problem domain and reproducing them using operators borrowed 
from natural genetics. This process leads to the evolution of populations of individuals that are better suited to their environment 
than the individuals that they were created from, just as in natural adaptation. 
The formal codification of a solution is called its ``genome'' (or genotype) while its actual behaviour 
is the ``phenotype''. Fitness is evaluated on the phenotype\footnote{There is a slight difference among the concept 
of fitness in the biological and evolutionary algorithms domains. In the former, fitness is associated to a phenotype 
by measuring the relative number of offspring it has generated; in the latter, on the contrary, an absolute value of 
fitness is associated to each phenotype and determines the relative number of offspring it will generate}.  
Evolutionary calculations often show the phenomenon of the ``code-bloat'' namely a major growth of the genome size
occurring without a significant improvement in fitness \cite{luke}. It manifests itself by the presence of regions 
of code that do nothing and can be mutated or removed without affecting the fitness. We will call these regions 
``non-coding''; a popular name for them is also ``introns'', in analogy with those portions in real genes that are 
not translated into sequences of aminoacids. We will call ``coding'' the active regions; they are often called 
``exons''. 
In the context of evolutionary algorithms, code-bloat constitutes a major problem; indeed it can lead to 
memory or computational time exhaustion before an optimal algorithm has been obtained. To avoid this phenomenon 
it is usually necessary to introduce a selective disadvantage against larger genomes; however, the selective
disadvantage has to be carefully tuned, since a bad choice can drastically slow down the evolution. 
Many hypotheses have been formulated to explain this phenomenon (specially in the framework
of ``genetic programming'') \cite{luke}:
``hitchhiking'', ``defence against crossover'',  ``removal bias'' and ``fitness caused diffusion''. 
The first two impute code-bloat to crossover. Indeed, according to the ``hitchhiking'' hypothesis, non-coding parts
attached to ``important'' parts of code are likely to be propagated in the genome by crossover. 
On the other hand, for the ``defence against crossover'' hypothesis, algorithms with large non-coding fraction
are selected since they have a smaller probability that their coding parts are disrupted by crossover.  
The ``removal bias'' hypothesis ascribes code-bloat to the fact that a deletion involving a non-coding part
will most probably become disadvantageous if it is larger of the non-coding part itself. On the other hand
an insertion in a non-coding part will always be neutral whatever be the size of the inserted code. So, selection 
will suppress large deletions but not large insertions, creating a bias in favour of the latter.  
Finally, up to the ``fitness caused diffusion'' hypothesis, the number of large-size highly-fit programs
is much larger than the small-size ones. So code-bloat can be described as the evolutionary model going toward 
its ergodic equilibrium. 

While the objects of evolution are very different in the biological and algorithm context, 
the concepts of survival, reproduction and selection are very similar so that the name of 
``artificial life'' is given to these ``in silico'' simulations of life \cite{fogel}.
This analogy makes plausible that some general features can be common to both systems.
In particular, the two phenomena of the C-value enigma and code bloat appear
surprisingly similar: the (genetic or algorithmic) code during its evolution devotes a large part of
itself to do, apparently, nothing. With the idea that the evolution of large non-coding parts is a  
feature of evolutionary models ``tout-court'', we decided to study the relations among fitness and coding/non-coding
ratio for different values of mutation and code growth rates in a particular evolutionary algorithm.
The advantage of working with evolutionary algorithms is that we can exactly associate a fitness value 
to an individual's genotype and we can clearly 
and unambiguously distinguish its non-coding by its coding parts. We will use Turing machines (TM) to encode
our algorithms, identifying the set of their internal states with the genome\footnote{For our aims using TMs is more
convenient than using tree based programs as in genetic programming; indeed, we can define independent rates of mutation
and code growth, whereas, in genetic programming, mutation also affects genome size.}. 
Specifically, we start with an initial population of trivial $1$-state Turing
machines and let them evolve for many generations. At each generation every
Turing machine undergoes three processes: mutation (with a rate $\pmm$),
states-increase (with a rate $\pii$) and selection/reproduction\footnote{We discarded crossover in this step to keep 
the model as simple as possible.} (measuring the relative fitness of the 
machines according with a properly specified task). 
Each Turing machine state is characterised by two triplets.  
The triplets of the final population are divided into non-coding and coding triplets. We study
how the fitness and the coding/non-coding ratio vary changing the values of the mutation ($\pmm$) and 
states-increase ($\pii$) rates.

%%%%%%%%%%%%%%%%%%%%%%%%%%%%%%%%%%%%%%%%%%%%%%%%%%%%%%%%%%%%%%
\section{What is a Turing machine\label{turing}}        
Turing machines are very simple symbol-manipulating devices which can be used to encode any feasible algorithm. 
They were invented in 1936 by Alan Turing \cite{Turing} and used as abstract tools to investigate the problem
of functions computability.

\begin{figure}[h]
\setlength{\unitlength}{0.9mm}
%\begin{picture}(200,29)(-6,-5)
%\multiput(0,0)(0,10){2}{\line(1,0){165}}
%\multiput(2,0)(10,0){17}{\line(0,1){10}}
%\multiput(0,0)(0,10){2}{\dottedline{1.4}(0,0)(-5,0)}
%\multiput(165,0)(0,10){2}{\dottedline{1.4}(0,0)(5,0)}
%\multiput(6,3.5)(10,0){3}{0}
%\put(36,3.5){1}\put(46,3.5){1}\put(56,3.5){0}\put(66,3.5){1}\put(76,3.5){0}
%\put(86,3.5){1}\put(96,3.5){0}\put(106,3.5){1}\put(116,3.5){1}\put(126,3.5){1}
%\multiput(136,3.5)(10,0){3}{0}
%\put(57,25){\drawline(-7,0)(-7,-5)(0,-14)(7,-5)(7,0)}
%\put(54,20){$\mathbf{s}(t)$}
%\put(54,-5){$k(t)$}
%\end{picture}
%\includegraphics[bb=60 14 498 458,clip,scale=0.6]{grafico1.ps}
\includegraphics{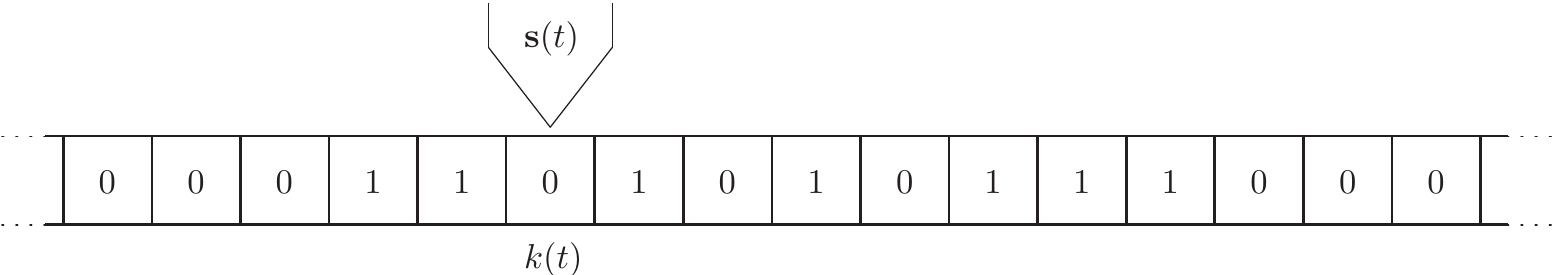}
\caption{Graphical representation of a Turing machine at time $t$, in the internal state $\mathbf{s}(t)$, 
located on the $k(t)$-th cell of a infinite tape.}
\end{figure}

In the following we give a description of Turing machines adapted to our purposes.
For a complete treatment of this subject we refer to  \cite{davis}.
Turing machine consists of a movable head acting on an infinite tape. The tape consists of discrete cells that can contain
a 0 or a 1 symbol. The head has a finite number of internal states.
At any time $t$ the head is in a given internal state $\mathbf{s}(t)$ and it is located upon a single cell $k(t)$.
It reads the symbol stored inside the cell and, according to its internal state and the symbol read, performs
three actions: 
\begin{enumerate}
\item ``write'':  writes a new symbol on the $k(t)$ cell, 
\item ``move'': moves one cell on the right or on the left or stay still ($k(t) \mapsto k(t+1)$),
\item ``call'': changes its internal state to a new state ($\mathbf{s}(t) \mapsto \mathbf{s}(t+1)$).
\end{enumerate}       
Accordingly, a state can be specified by two triplets ``write-move-call'' listing the actions to undertake 
after reading respectively a $0$ or $1$ symbol. The following is an example of a possible state
$$
\begin{array}{|c|c|}
\hline
 & \mathbf{r} \\ \hline
 0 & 1-\mbox{Right}-{\bf 3} \\ \hline
 1 & 0-\mbox{Left}-{\bf 7} \\ \hline
\end{array}
$$
so that, if the head is in the state $\mathbf{s}(t)=\mathbf{r}$ and  reads $0$, it writes a $1$, moves one cell right and goes in the state $\mathbf{3}$
while if it reads 1, writes $0$, moves left and goes in the state $\mathbf{7}$.   
There is a distinguished state that causes the machine to halt (``\textbf{Halt} state'').
The initial tape ($t=0$) specifies the input data for the algorithm encoded by the Turing machine while the final tape 
(i.e. the one obtained after the machine has halted) gives the output. The head of the machine is initially in the state 
$\mathbf{s}(0)=\mathbf{1}$ and it is conventionally located upon the cell containing the leftmost $1$. 
 
We give here a working example of a Turing machine that performs the sum of two numbers
\setlength{\unitlength}{1mm}
$$
\begin{array}{|c|c|c|c|}
\hline \raisebox{-2mm}[7mm][5mm]{read} \raisebox{3mm}[7mm][5mm]{\bf state} 
%\begin{picture}(15,10)\put(12,3){\line(-1,1){9}} \put(1,3){\mbox{read}}\put(7,9){\mbox{\bf %state}}\end{picture}
& {\bf 1} & {\bf 2} & {\bf 3}  \\
\hline 0 & 1-\mbox{Right}-\bf 2 & 0-\mbox{Left}-\bf 3 & \_-\_-\_  \\
\hline 1 & 1-\mbox{Right}-\bf 1 & 1-\mbox{Right}-\bf 2 & 0-\_-\mbox{\bf Halt}  \\ \hline
\end{array}
$$
where the underscore symbol means that the corresponding entry can be arbitrarily chosen without affecting the algorithm.
A positive integer number $n$ is represented on the tape by a contiguous string of $n$-ones preceded and followed by a zero digit, 
for example $\ldots 01110\ldots $ represents the number three. Then the sum of $3+2$ is performed by the given machine 
in the following steps:
$$
\begin{array}{cccccccccccccccccc}
& {\bf 1} &   &   &   &   &   &   &  & &&{\bf 1} & & &  &   &   &       \\[-2mm]
\ldots 0& 1 & 1 & 1 & 0 & 1 & 1 & 0 \ldots & \xrightarrow{\mbox{\small 1-Right-\bf 1}} & \ldots 0& 1 & 1 & 1 & 0 & 1 & 1 & 0 \ldots & 
\xrightarrow{\mbox{\small 1-Right-\bf 1}} \\[2mm]
  &   &  & {\bf 1}  &  & &   &   &  & &  & &   & {\bf 1}  &  &   &      &  \\[-2mm]
\ldots 0&1 & 1 & 1 & 0 & 1 & 1 & 0\dots & \xrightarrow{\mbox{\small 1-Right-\bf 1}} & \ldots 0 & 1 & 1 & 1 & 0 & 1 & 1 & 0 \ldots &
\xrightarrow{\mbox{\small 1-Right-\bf 2}} \\[2mm]
  &&   &   &   &{\bf  2} &   &&   &  & &    &   &   &   & {\bf 2}  &  &    \\[-2mm]
\ldots 0& 1 & 1 & 1 & 1 & 1 & 1 & 0  \dots & \xrightarrow{\mbox{\small 1-Right-\bf 2}} & \ldots 0& 1 & 1 & 1 & 1 & 1 & 1 & 0 \ldots &
\xrightarrow{\mbox{\small 1-Right-\bf 2}} \\[2mm]
&  &   &   &   &   &   & {\bf 2}\,\phantom{\ldots} && & &   &   &   &   & {\bf 3}  &  &    \\[-2mm]
\ldots 0&1 & 1 & 1 & 1 & 1 & 1 & 0 \ldots & \xrightarrow{\mbox{\small 0-Left-\bf 3}} & \ldots 0& 1 & 1 & 1 & 1 & 1 & 1 & 0 \ldots  & 
\xrightarrow{\mbox{\small 0-\_-\bf Halt}} \\[2mm]
  &   & &&  &   &  &   &   & & & & & & &  & &\\[-2mm]
\ldots 0&1 & 1 & 1 & 1 & 1 & 0 & 0 \ldots && &  & & & & & & &
\end{array}
$$
In the paradigm of artificial life, a Turing machine is an organism; its set of states is the genotype
and the output tape is the phenotype. 
In this frame it is natural to classify as ``non-coding'' those triplets that are never used by the TM, as the 
third state triplet ($\_-\_-\_$) corresponding to the read value 0 in the above machine. Notice that whatever input tape is provided, this triplet is never used. 
In general, the most common case is that the distinction
between coding and non-coding triplets depends on the specific choice of the input tape. 
In this sense the input tape can be interpreted as an extremely powerful epigenetic conditioning that influences 
the activation of the various triplets of the TM. 

The finite state machines used in the context of evolutionary programming (L. Fogel, 1962 \cite{fogel}) 
are a subclass of Turing machines with right only movement. We will not make use of them but cite them as
the first appearance of similar devices in the field of artificial life.

\section{The model\label{model}}
Since we want to perform computer simulations, we need to use a tape of finite length that we fix to 300 cells.
Conventionally, our machines always start from the leftmost cell whatever its content is. 
If the machine runs out of the tape
(both on the left or on the right) it is halted. 
Since it is quite easy to generate machines that run forever, we also need to fix a maximum number of time steps, 
therefore we choose to force halting the machine if it reaches 4000 steps.

We begin with a population of $300$ $1$-state TM of the following form 
$$
\begin{array}{|c|c|}
\hline
 &\bf 1 \\ \hline
 0 & 0-\mbox{Still}-\mbox{\bf Halt} \\ \hline
 1 & 1-\mbox{Still}-\mbox{\bf Halt} \\ \hline
\end{array}
$$
and let them evolve for $200000$ generations.
At each generation every TM undergoes the following three processes (in this order):
\begin{enumerate}
\item states-increase,
\item mutation,
\item selection and reproduction.
\end{enumerate}
\paragraph{State-increase.} In this phase, with a probability $\pii$, the TM passes from $\mathbf N$ to $\mathbf N+\mathbf 1$ states by the addition of  the further state
$$
\begin{array}{|c|c|}
\hline & \mathbf N+\bf 1\\
\hline 0 & 0-\mbox{Still}-\mbox{\bf Halt}\\
\hline 1 & 1-\mbox{Still}-\mbox{\bf Halt}\\ \hline
\end{array}
$$
As it will become clear from the definition of the mutation process, this state will be initially non-coding since it cannot be called by any other state. The only way it can be activated is if a mutation  
in a coding state changes the state call to $\mathbf N+\mathbf 1$. Notice that, when called, this particular state does not change the tape and halts the machine.
Consequently the activation of this state is mainly harmful or neutral and it can be advantageous only in exceptional cases, so that the TM can benefit from the added states only if they are mutated before their activation.    

\paragraph{Mutation.}
During mutation, all entries of each state of the TM are randomly changed with probability $\pmm$. The new entry is randomly chosen 
among all corresponding permitted values excluding the original one. 
The permitted values are: 
\begin{itemize}
\item 0 or 1 for the ``write'' entries;
\item Right, Still, Left for the ``move'' entries;
\item The Halt state or an integer from {\bf 1} to the number of states $\mathbf N$ of the machine for the ``call'' entries.
\end{itemize}
This mechanism of mutation is reminiscent of the biological point mutation. We have not implemented other biological mechanisms like traslocation, inversion, deletion, etc. 

\paragraph{Selection and reproduction.}
In the selection and reproduction phase a new population is created from the actual one (old population).
The number of offspring of a TM is determined by its ``fitness'' and, to a minor extent, by chance.
The fitness of a TM is a function that measures how well the output tape of the machine reproduces a given ``goal'' tape
starting from a prescribed input tape. We compute it in the following way. The fitness is initially set to zero. 
Then the output tape and the goal tape are compared cell by cell. The fitness is increased by one for any $1$ on the
output tape that has a matching $1$ on the goal tape and it is decreased by 3 for any $1$ on the output tape
that matches a $0$ on the goal tape. 
    
As a selection process, we use what in the field of evolutionary algorithms is known as ``tournament selection of size 2''. 
Namely two TMs are randomly extracted from the old population, they run on the input tape and a fitness value is
assigned to them according to their output tapes. The fitness values are compared and the machine which scores higher creates two copies 
of itself in the new population, while the other is eliminated (asexual reproduction). If the fitness values are equal, each TM creates
a copy of itself in the new population.
The two TMs that were chosen for the tournament are eliminated from the old population and the process restarts until the exhaustion 
of the old population.

For our aims, this selection mechanism has many advantages. Since increase in the fitness is quite rare we are strongly interested 
in the survival of the best TM, that is automatically granted by the selection mechanism. 
Moreover, having a very small population (due to computational time reasons) we would like to maintain a maximal ``biodiversity''. 
On the other hand selection implies that TMs with higher fitness have to generate an higher number of offspring, so that they will eventually colonize all the population decreasing the biodiversity. Our selection mechanism 
ensures that this colonization does not happen too fast. Indeed, the expected number of offspring for the TMs 
belonging to the best fitness group (those Turing machines that share the best fitness) varies from $1$ to $2$,
depending on the size of this group. The expected number of offspring will be $2$ only when there is a single
TM in the group and it will progressively decreases to $1$ when the group size increases. Finally it will be 
exactly one when the best fitness group coincides with the whole population.     
Another phenomenon limiting biodiversity is genetic drift. In our case such phenomenon is completely absent, 
since the tournament preserves both TMs if they score equal\footnote{Observe that two 
TMs can be very different and share the same score.}. 
This feature makes this selection procedure also computationally fast, since in the case of an even result
the code needs to do nothing.

\section{The simulations}
In this section we discuss the various choices of parameters adopted for computer simulations with our model.
The initial tape was permanently fixed to contain only zeroes. Of course, many other choices are possible, as giving a
completely different initial tape or varying it with generations. As we said in section~\ref{turing}, the input tape
can be interpreted as an epigenetic conditioning. Since we want to keep our model as simple as possible, we decided
to keep the input tape fixed. Since we use the symbol $1$ to measure the fitness, a tape made entirely of $0$ is the most 
convenient choice. We will define non-coding triplets relatively to this choice of input tape. 
That is a triplet of a TM will be called non-coding if it is never executed 
when the TM runs on the input tape made of $300$ zeroes.
This implies that the values of a non-coding triplet 
can be arbitrarily changed without affecting the corresponding output tape.

We performed many simulations with different values of the rates $\pii, \ \pmm$, two choices for  the ``goal'' tape and
ten choices for the seed of the random number generator. 

The states-increase rate has been chosen in the following set of values 
\begin{equation}
\pii\in \left\{ \frac{1}{10800}\,;\frac{1}{6000}\,;\frac{1}{3333}\,;\frac{1}{1852}\,;\frac{1}{1029}\,;\frac{1}{572}\,;
\frac{1}{318}\,;\frac{1}{176}\,;\frac{1}{98}\,;\frac{1}{54}\,;\frac{1}{30} \right\}\,.
\end{equation}
These values were generated starting from the smallest one and requiring an approximate $\frac59$ ratio between consecutive numbers. 
As resulted from some trials, this particular ratio was seen to be the optimal one.  

The mutation rate takes the following values
\begin{equation}
\pmm\in \left\{ \frac1{20360}\,; \frac1{12339}\,; \frac1{7478}\,; \frac1{4532}\,; \frac1{2747}\,; \frac1{1665}\,; \frac1{1009}\,; 
\frac1{611}\,; \frac1{371}\,; \frac1{225}\,;\frac1{136}\,; \frac1{83}\,; \frac1{50}\right\}
\end{equation}
constructed in the same way as $\pii$ but with an approximate ratio $0.6$.

The goal tapes are chosen according to the criterion of providing two difficult and qualitatively different tasks for a TM; in this sense
the distribution of the ones on the goal tape has to be extremely non-regular since a periodic distribution
is a very easy task for a TM. 

We decided to use a goal tape with ones on the cell positions corresponding to prime 
numbers (with $1$ included for convenience) and zeroes elsewhere:
$$
\begin{array}{l}
1110101000.1010001010.0010000010.1000001000.1010001000.0010000010.1000001000.1010000010.\\
0010000010.0000001000.1010001010.0010000000.0000001000.1000001010.0000000010.1000001000.\\
0010001000.0010000010.1000000000.1010001010.0000000000.1000000000.0010001010.0010000010.\\
1000000000.1000001000.0010000010.1000001000.1010000000.0010000000.
\end{array}
$$
In the previous expression we inserted a dot every ten cells to facilitate the reading.
Our second goal tape is given by the binary expression of the decimal part of $\pi$, namely $(\pi-3)_{\text{bin}}$:
$$
\begin{array}{l}
0010010000.1111110110.1010100010.0010000101.1010001100.0010001101.0011000100.1100011001.\\
1000101000.1011100000.0011011100.0001110011.0100010010.1001000000.1001001110.0000100010.\\
0010100110.0111110011.0001110100.0000001000.0010111011.1110101001.1000111011.0001001110.\\
0110110010.0010010100.0101001010.0000100001.1110011000.1110001101.
\end{array}
$$
Notice that while for prime numbers the ones become progressively rarer so that the task becomes progressively more difficult, in the case of the digits of $\pi$ the ones are more or less equally distributed. 
Another difference is that prime numbers are always odd (with the exception of 2) so that in the goal tape
two ones are separated by at least one zero. On the contrary, the digits of $\pi$ can form clusters of ones of arbitrary length.

According to our definition, the maximal possible value for the fitness is 63 for the prime numbers and 125 for the digits of $\pi$.

The program for the simulation has been written in C and we used the native random number generator. We tested that its randomness 
is suitable for our purposes.

We provide an example of a TM obtained after $20000$ generations, with states-increase rate \mbox{$\pii=1/3333$},  
mutation rate $\pmm=1/1009$ and prime numbers task.

\begin{center}
\begin{tabular}{|c|c|c|c|c|c|c|c|c|}
\hline  &\bf 1 &\bf 2 &\bf 3 &\bf 4 &\bf 5 &\bf 6 &\bf 7 &\bf 8 \\
\hline 0 & {\red 1-Right-\bf 4} & {\red 1-Right-\bf 5} & \red{1-Right-\bf 6} & {\red 1-Right-\bf 3} & {\red 1-Right-\bf 7} & {\red 0-Right-\bf 2}& \red{1-Left-\bf 7}   & 0-Still-\bf 8\\
\hline 1 & 1-Still-\bf 3 & 1-Still-\bf 2 & 0-Still-\bf 3 & 0-Still-\bf 5  & 0-Left-\bf 4  & 1-Left-\bf 1 & {\red 0-Left-\bf Halt} & 0-Right-\bf 2 \\ \hline
\end{tabular}

\bigskip
\begin{tabular}{ccccccccccccccccccc}
\bf 1 &   &   &   &   &   &   &  & & & \bf 4 &   &   &   &   &   & &\\[-2mm]
0 & 0 & 0 & 0 & 0 & 0 & 0 & 0 \dots & $\xrightarrow{\mbox{\small 1-Right-\bf 4}}$ & 1 & 0 & 0 & 0 & 0 & 0 & 0 & 0 \dots & $\xrightarrow{\mbox{\small 1-Right-\bf 3}}$ \\[2mm]
  &   & \bf 3 &   &   &   &   &  & &   &   &   &\bf 6 &   &   &   & & \\[-2mm]
1 & 1 & 0 & 0 & 0 & 0 & 0 & 0 \dots & $\xrightarrow{\mbox{\small 1-Right-\bf 6}}$ & 1 & 1 & 1 & 0 & 0 & 0 & 0 & 0 \dots & 
$\xrightarrow{\mbox{\small 0-Right-\bf 2}}$ \\[2mm]
  &   &   &   & \bf 2 &   &   &  & &    &   &   &   &   & \bf 5 &   &  & \\[-2mm]
1 & 1 & 1 & 0 & 0 & 0 & 0 & 0 \dots & $\xrightarrow{\mbox{\small 1-Right-\bf 5}}$ & 1 & 1 & 1 & 0 & 1 & 0 & 0 & 0 \dots &
$\xrightarrow{\mbox{\small 1-Right-\bf 7}}$ \\[2mm]
  &   &   &   &   &   & \bf 7 & & &   &   &   &   &   & \bf 7 &   & & \\[-2mm]
 1 & 1 & 1 & 0 & 1 & 1 & 0 & 0 \dots & $\xrightarrow{\mbox{\small 1-Left-\bf 7}}$ & 1 & 1 & 1 & 0 & 1 & 1 & 1 & 0 \dots
  & $\xrightarrow{\mbox{\small 0-Left-\bf Halt}}$ \\[2mm]
  &   &   &   & \bf H &   &   & & & & & & & & & & &\\[-0.5mm]
1 & 1 & 1 & 0 & 1 & 0 & 1 & 0 \dots & & & & & & & & & & 
\end{tabular}
\end{center}

We observe that this TM reaches a fitness of 5 with 8 states. The 8 coding triplets are written in red while the non-coding ones
are written in black. Accordingly, the coding triplets amount to $50\%$ of the total triplets. 
We notice also that this machine writes a 1 in the sixth cell only to come back later and cancel it. So this is clearly 
not the most economic TM with the same performance. As we will see this is a typical feature of the TMs obtained
through our model, for certain choices $\pii,\ \pmm$.

%%%%%%%%%%%%%%%%%%%%%%%%%%%%%%%%%%%%%%%%%%%%%%%%%%%%%%%%%%%%%%%%%%%%%%%%%%%%%%%%
\section{Results} \label{results}

\begin{figure}[!h]
%\setlength{\unitlength}{1mm}
%\begin{picture}(150,163)(0,-70)
%\put(8,0){\includegraphics[bb=60 14 498 458,clip,scale=0.6]{Primi3d.eps}}
%\put(100,20){\includegraphics[bb=94 110 498 478,clip,scale=0.6]{Primicontour.eps}}
%\put(55,5){\color{white}\rule{13mm}{5mm}}\put(5,17){\color{white}\rule{13mm}{5mm}}
%\put(58,8){$\log_{10} \pmm$}\put(8,20){$\log_{10} \pii$}
%\put(130,18){$\log_{10} \pmm$}\put(93,50){$\log_{10} \pii$}
%\put(6,60){\begin{rotate}{90}fitness\end{rotate}}
%\put(15,-60){\includegraphics[bb=100 110 498 408,clip,scale=0.6]{Primi_p_i.eps}}
%\put(100,-60){\includegraphics[bb=100 110 498 408,clip,scale=0.6]{Primi_p_m.eps}}
%\put(45,-65){$\log_{10} \pii$}\put(130,-65){$\log_{10} \pmm$}
%\put(12,-27){\begin{rotate}{90}fitness\end{rotate}}\put(97,-27){\begin{rotate}{90}fitness\end{rotate}}
%\put(48,4){(a)}\put(133,4){(b)}\put(48,-72){(c)}\put(133,-72){(d)}
%\end{picture}
%\includegraphics[bb=90 290 558 760]{grafico2.pdf}
\includegraphics{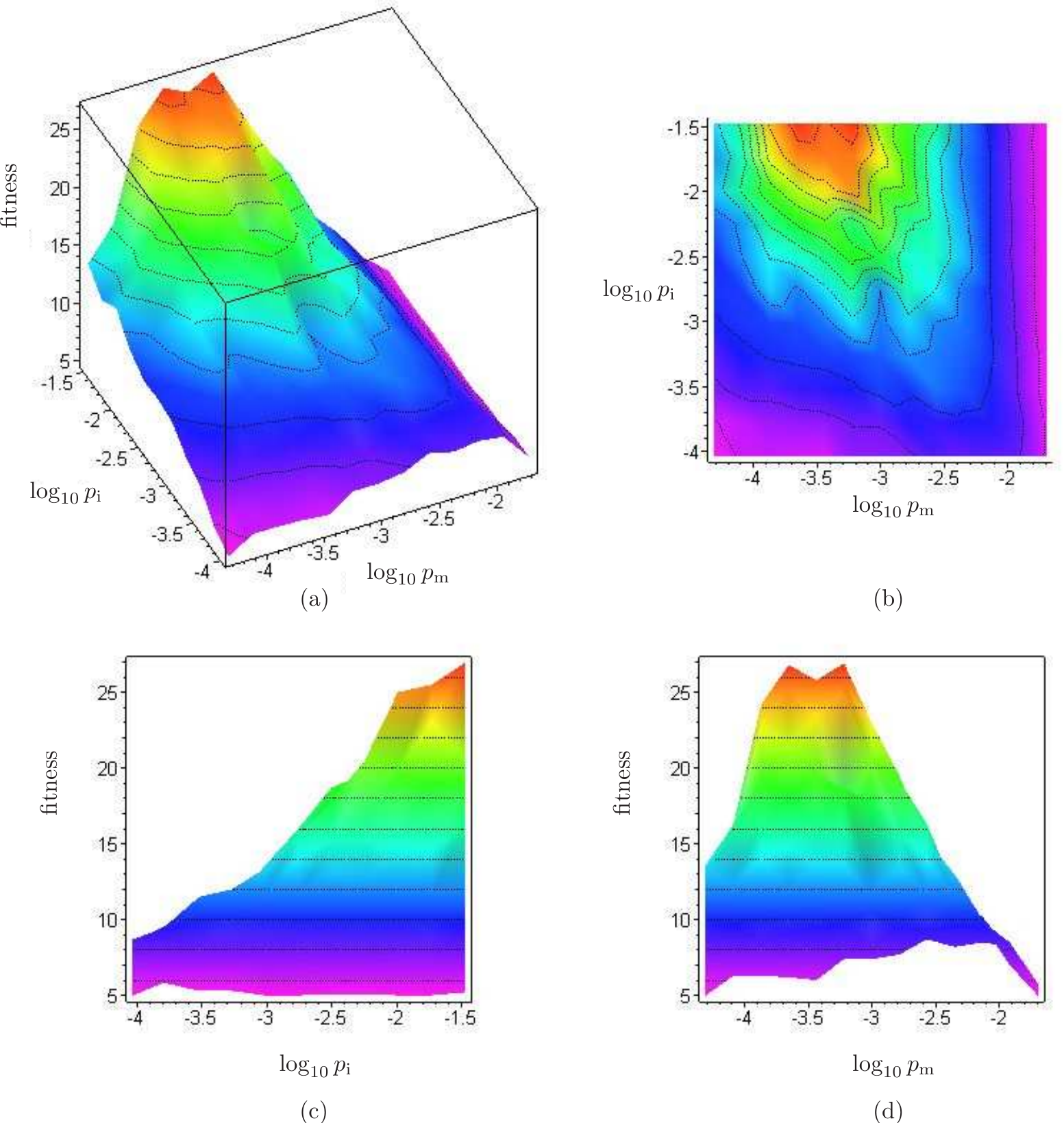}
\caption{\label{plotprimi}For the case of prime numbers, in (a) we show the 3D plot of the best fitness value 
in the population, averaged on the ten different seeds, as function of the states-increase rate $\pii$ and 
of the mutation rate $\pmm$. The three orthogonal projections of (a) are also shown.}
\end{figure}

\begin{figure}[!h]
%\setlength{\unitlength}{1mm}
%\begin{picture}(150,163)(0,-70)
%\put(8,0){\includegraphics[bb=60 14 498 458,clip,scale=0.6]{Pi3d.eps}}
%\put(100,20){\includegraphics[bb=94 110 498 478,clip,scale=0.6]{Picontour.eps}}
%\put(55,5){\color{white}\rule{13mm}{5mm}}\put(5,17){\color{white}\rule{13mm}{5mm}}
%\put(58,8){$\log_{10} \pmm$}\put(8,20){$\log_{10} \pii$}
%\put(130,18){$\log_{10} \pmm$}\put(93,50){$\log_{10} \pii$}
%\put(6,60){\begin{rotate}{90}fitness\end{rotate}}
%\put(15,-60){\includegraphics[bb=100 110 498 408,clip,scale=0.6]{Pi_p_i.eps}}
%\put(100,-60){\includegraphics[bb=100 110 498 408,clip,scale=0.6]{Pi_p_m.eps}}
%\put(45,-65){$\log_{10} \pii$}\put(130,-65){$\log_{10} \pmm$}
%\put(12,-27){\begin{rotate}{90}fitness\end{rotate}}\put(97,-27){\begin{rotate}%{90}fitness\end{rotate}}
%\put(48,4){(a)}\put(133,4){(b)}\put(48,-72){(c)}\put(133,-72){(d)}
%\end{picture}
%\includegraphics[bb=90 290 558 760]{grafico3.pdf}
\includegraphics{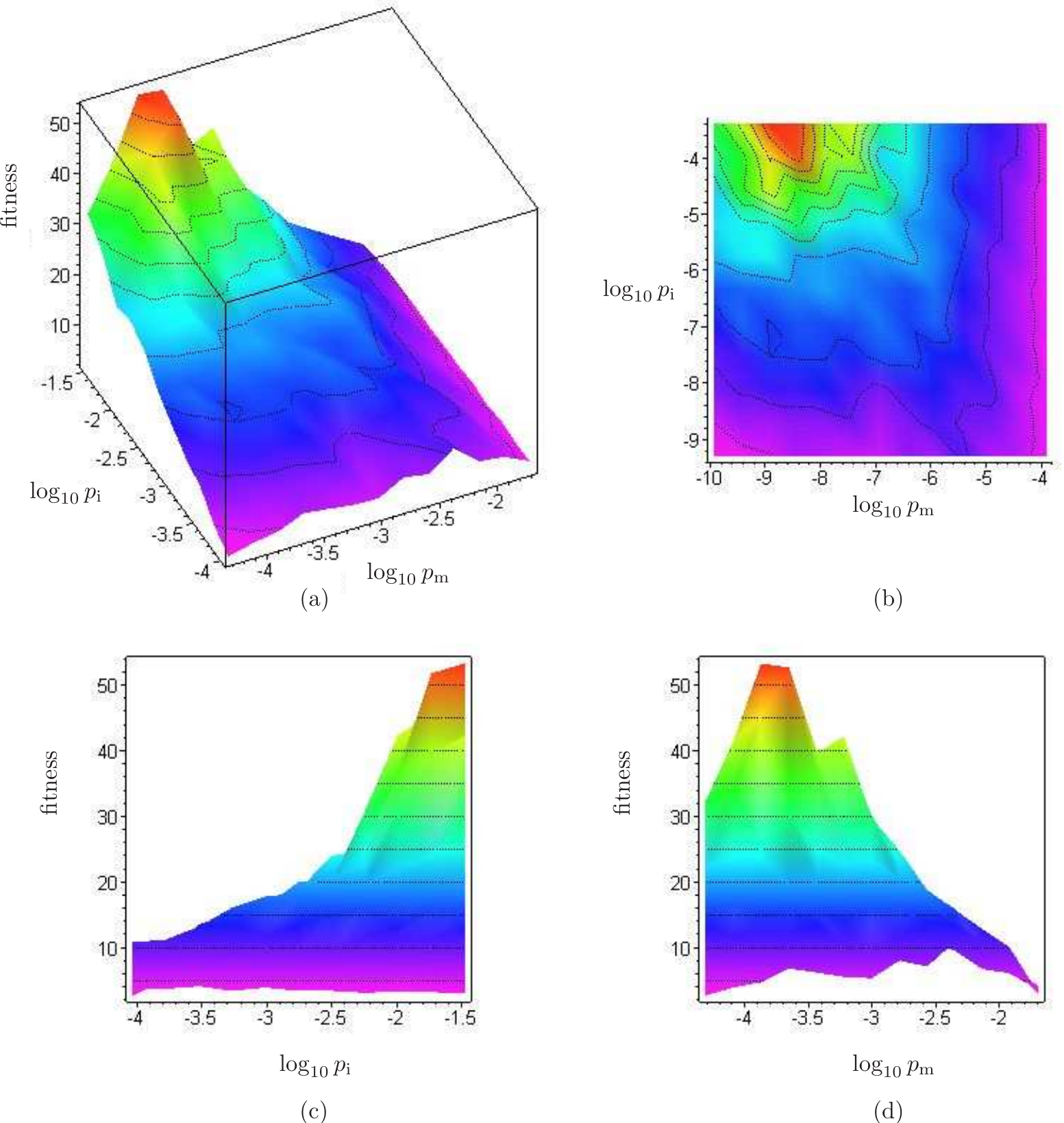}
\caption{\label{plotpi}For the case of digits of $\pi$, in (a) we show the 3D plot of the best fitness value 
in the population, averaged on the ten different seeds, as function of the states-increase rate $\pii$ and
of the mutation rate $\pmm$. The three orthogonal projections of (a) are also shown.}
\end{figure}

\begin{figure}[!h]
\hspace*{15mm}
\includegraphics[viewport=50 70 570 560,scale=0.5]{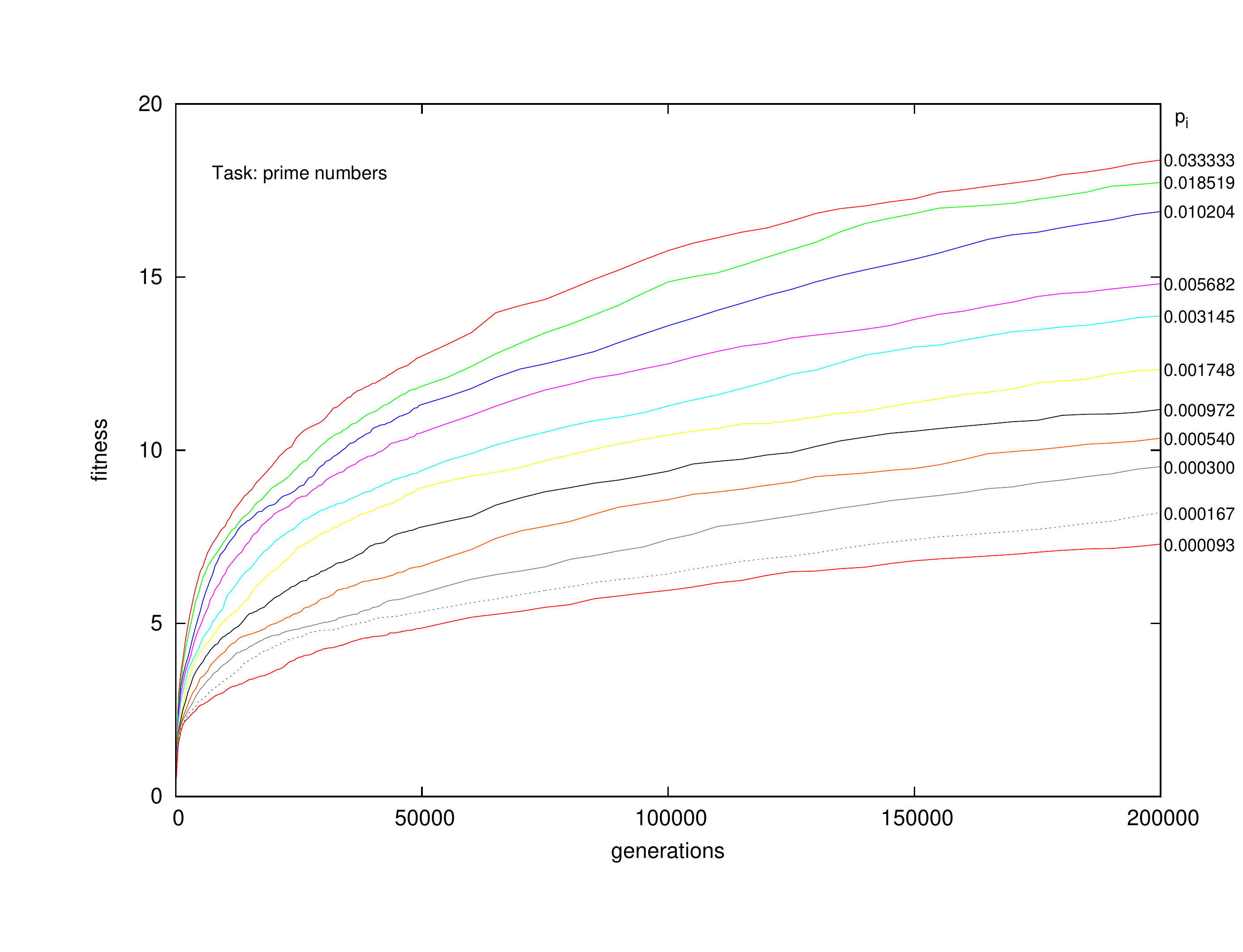}
\caption{\label{storia_primes}Evolution of the population best fitness with generations, averaged on the seeds 
and the mutation probabilities.}

\vspace{5mm}
\hspace*{15mm}
\includegraphics[viewport=50 70 570 560,scale=0.5]{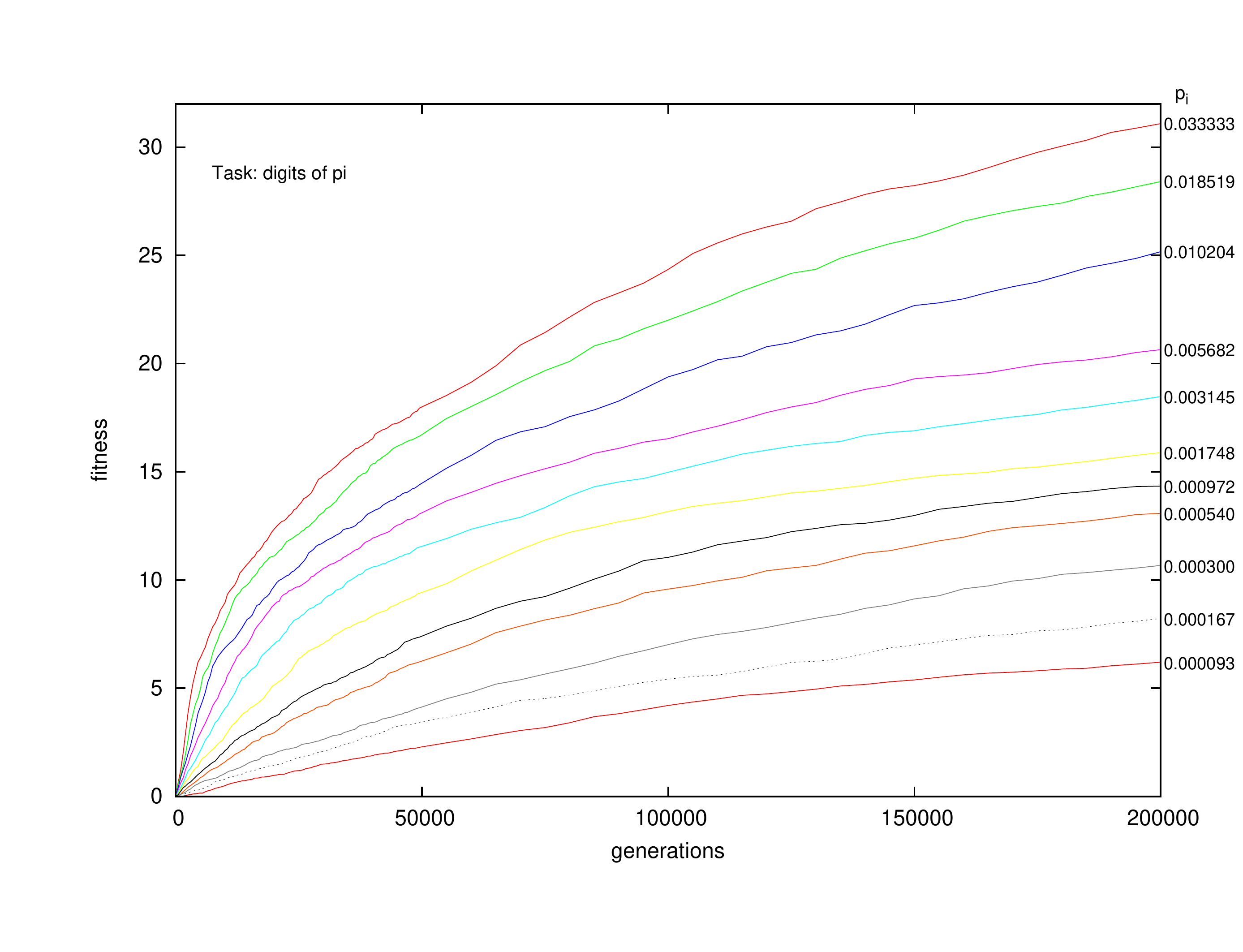}
\caption{\label{storia_pi}Evolution of the population best fitness with generations, averaged on the 
seeds and the mutation probabilities.}
\end{figure}

We start examining the best fitness reached by the population of Turing machines, as shown in figures~\ref{plotprimi} and \ref{plotpi}.
The most evident effect is the mainly monotonic growth of fitness with the states-increase rate $\pii$,
(see in particular the figures \ref{plotprimi}.c and \ref{plotpi}.c), occurring at almost all values of the mutation rate.
This effect is particularly interesting in combination with the observation that the total number of states of a TM grows 
approximately proportional to $\pii$ (actually, it is approximately given by the number of generations times the 
states-increase rate; see later, eq.~\ref{nstates}) therefore we state that populations of machines with larger 
number of states are those where a higher best fitness can be obtained. 
Similarly, the figures~\ref{storia_primes} and \ref{storia_pi} indicate that populations of machines with a larger number of states, 
i.e. obtained with a larger value of $\pii$, reach a better fitness at all times, namely after any number of generations. 
Indeed, for both tasks, the various curves are ordered according to increasing values of $\pii$ and
they never intersect. These are strong indications that, in our model, the rate of evolution of a population of Turing machines
is directly related to the rate of increase of the number of its states.

Since the maximum values of fitness are obtained at the largest examined values of $\pii$, 
from figures~\ref{plotprimi} and \ref{plotpi} we cannot state if they are true maxima (either absolute or relative) or possibly if a 
saturation will occur. 
To understand this would require to obtain new data with values of $\pii$ larger than $\frac{1}{30}$ but this would 
largely increase 
the needed computational time. Indeed, a run with $\pii=\frac{1}{30}$ requires about 100 times 
the computational time needed for a run with $\pii=\frac{1}{10800}$. This great difference is 
caused by the fact that the number of states of a TM grows approximately as the number of generations 
times the states-increase rate. 
In section~\ref{comparative} we will test for the existence of a maximum at large values of $\pii$ by performing a simplified simulation.

Another way to test the role of the states-increase rate on fitness is to look at the final number 
of states. Indeed, since the states-increase 
procedure alone does not affect the fitness, one could expect that the number of states is not subject to
selection. On the other hand, if machines with more states evolve faster, one should expect that the number
of states has to be positively selected. We stress the fact that this selection mechanism is completely indirect,
in the sense that there is a greater probability that a TM with more states develops an advantageous mutation.
In the following table~\ref{tabN} we report the observed final number of states $\bar{N}_{\text{obs}}$ averaged on the population, 
the mutation rate and the seeds, for all values of the states-increase rate. 
We also report the number of states $N_{\text{exp}}$, expected under the assumption that it is not a selected character 
so that it is determined only by chance:
\begin{equation}\label{nstates}
N_{\text{exp}}=200000\ \pii +1\,.
\end{equation}
Finally we also indicate the relative difference among the two values, 
$$\%~\mbox{diff}=100 \frac{\bar{N}_{\text{obs}}- N_{\text{exp}}}{N_{\text{exp}}} \,.$$

From table \ref{tabN}, we observe for both tasks that $\bar{N}_{\text{obs}}$ is significantly larger than 
$N_{\text{exp}}$, for $\pii$ small. 
This indicates that, in such cases, there is a positive selection on the number of states\footnote{Simultaneous 
mutations in a given TM are allowed. This means that the growth in the number of states cannot act as a
shield against harmful mutations.}.   
 
Moreover, it is evident that the relative difference between $\bar{N}_{\text{obs}}$ and 
$N_{\text{exp}}$  decreases progressively as $\pii$ increases.
This decrease is an expected phenomenon, since also the relative standard deviation of the distribution of $N_{\text{exp}}$ 
decreases while $\pii$ increases:
$$
\frac{\sigma}{N_{\text{exp}}} \simeq \frac1{\sqrt{200000\  \pii}} 
$$ 

Finally we observe that for the primes task the relative differences are larger than those of the digits of $\pi$ task.
This effect is probably due to the fact that it is initially easier for a TM to increase its fitness for the primes
task than for the digits of $\pi$ one, as can be clearly seen from figures~\ref{storia_primes} and \ref{storia_pi};
this creates a strong selective bias towards TMs with more states for the first generations. This initial easiness
basically follows from the three consecutive ones at the very beginning of the primes goal tape.

\begin{table}[!h]
\caption{\label{tabN}For the two tasks, we report the observed and expected number of states and their relative difference.}

\vspace*{-4mm}
$$
\begin{array}{|c||c|c|c|c|c|c|c|c|c|c|c|c|} 
 \hline \rule{0mm}{4mm} \raisebox{-3mm}[0mm][0mm]{task}
&\pii  & \frac1{10800} & \frac1{6000} &\frac1{3333} &  \frac1{1852} &  \frac1{1029} & \frac1{572} & \frac1{318} &   \frac1{176} & \frac1{98} & \frac1{54} &    \frac1{30}\\[1mm] \cline{2-13} \rule{0mm}{4.5mm}
&N_{\text{exp}} &    19.5 &    34.3 &    61.0 &   109.0 &   195.4 &   350.7 &   629.9 &  1137.4 &  2041.8 &  3704.7 &  6667.7\\[1mm] \hline\hline \rule{0mm}{4.5mm}
\raisebox{-3mm}[0mm][0mm]{primes}&\bar{N}_{\text{obs}} &    23.5 &    37.9 &    64.2 &   113.5 &   199.4 &   355.1 &   633.0 &  1144.2 &  2045.0 &  3701.2 &  6670.1\\[1mm] \cline{2-13} \rule{0mm}{4.5mm}
&\%~ \mbox{diff} &    20.48 &    10.49 &     5.18 &     4.12 &     2.05 &     1.27 &     0.49 &     0.60 &     0.15 &    -0.09 &     0.04\\[1mm] \hline\hline \rule{0mm}{4mm}
\raisebox{-3mm}[0mm][0mm]{$\pi$}&\bar{N}_{\text{obs}} & 21.8 & 36.2 &    63.3 &   111.1 &   197.5 &   356.4 &   631.0 &  1134.6 &  2045.4 &  3710.0 &  6665.6\\[1mm] \cline{2-13} \rule{0mm}{4.5mm}
&\%~ \mbox{diff} &    11.66 &     5.48 &     3.71 &     1.96 &     1.10 &     1.63 &     0.17 &    -0.25 &     0.17 &     0.14 &    -0.03\\[1mm] \hline 
\end{array}$$
\end{table}

\begin{figure}[!h]
%\setlength{\unitlength}{1mm}
%\begin{picture}(150,85)
%\put(7,0){\includegraphics[bb=60 14 498 458,clip,scale=0.54]{esoni_primi.eps}}
%\put(91,0){\includegraphics[bb=60 14 498 458,clip,scale=0.54]{esoni_pi.eps}}
%\put(50,6){$\log_{10} p_m$}\put(8,16){$\log_{10} p_i$}
%\put(132,6){$\log_{10} p_m$}\put(90,16){$\log_{10} p_i$}
%\multiput(2,35)(86,0){2}{\begin{rotate}{90}number of coding triplets \end{rotate}}
%\put(40,0){(a)}\put(125,0){(b)}
%\end{picture}
\includegraphics{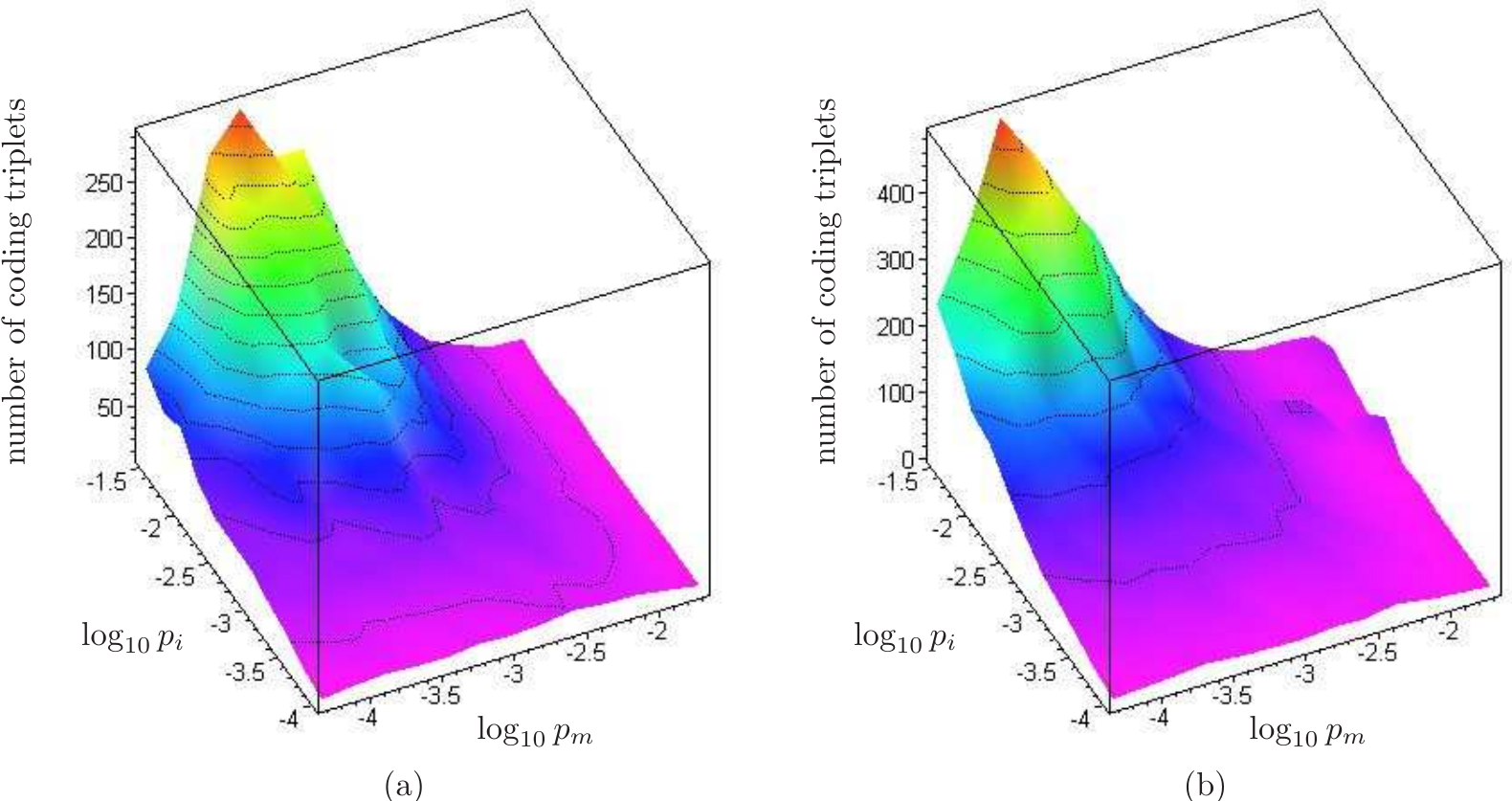}
\caption{\label{codingst}For the prime numbers task (a) and digits of $\pi$ task (b), we show the number of 
coding triplets averaged 
on the fittest machines in the population and on the seeds, for every $\pii,~ \pmm$.}
\end{figure}

In figure~\ref{codingst} we give a three dimensional view of the number of coding triplets averaged 
on the fittest machines in the population and on the seeds, for every $\pii,~ \pmm$. These plots closely resemble 
those for the
fitness~\ref{plotprimi}.a and \ref{plotpi}.a, showing a strict correlation between the two values. 
%In some amount, this 
%is consequence of the ``serial'' behaviour of the TM: they try to guess one digit ``1'' after the other. 
In particular, the maximum
fitness and maximum number of coding triplets occur at intermediate values of $\pmm$ while for both large and small ones
the fitness and number of coding triplets rapidly decrease. We give some explanations for this behaviour.

If the mutation rate is too small, it is clear that there is not enough variation between a machine and 
its offspring, for selection to work on.  

On the contrary, high mutation rates exert a limiting effect on the maximum number of coding triplets 
allowed to a TM.
Indeed a TM with many coding triplets will most probably undergo a mutation affecting its 
output tape. Since mutations in coding triplets are probably nefarious, such a TM is likely doomed to extinction.  
A rough estimate on the allowed maximum number of coding triplets can be done by assuming that their mutations are always 
lethal. Since each entry in the triplet has a probability $\pmm$ of mutating, the whole machine will undergo 
mutation with a rate $3 \nc\, \pmm$, where $\nc$ denotes the number of coding triplets.
Being $2$ the maximum number of children, the machine lineage will survive if it is more probable that only one child
is mutated than they are both.
This means that it must hold
$$
3 \nc\, \pmm< \frac12 \qquad \mbox{i.e.} \qquad \nc < \frac{1}{6 \pmm}
$$   
This estimate turns out to be very accurate for prime numbers task, and also, but with more exceptions
represented by the bumps in figure \ref{codingst}.b, for the digits
of $\pi$ task. So, the assumption that mutations occurring in coding triplets are always lethal is not too far from truth. 

For the sake of clarity, we report the fitness and the mean number of coding states for the maximum value of the states-increase rate
$\pii=\frac{1}{30}$, averaged on the seeds and on the TM with best fitness in the population.
$$\begin{array}{|c||c|c|c|c|c|c|c|c|c|c|c|c|c|c|}
\hline \rule{0mm}{4mm} \mbox{task}
&\pmm & \frac1{20360} & \frac1{12339} &\frac1{7478} &  \frac1{4532} &  \frac1{2747} & \frac1{1665} & \frac1{1009} &   \frac1{611} & \frac1{371} & \frac1{225} & \frac1{136}&\frac1{83}&\frac1{50}\\[1mm] \hline\hline \rule{0mm}{4.5mm}
\raisebox{-3mm}[0mm][0mm]{primes}&\mbox{fitness}&13.6   & 16.3   & 24.2   &26.8   &25.8   &26.9   & 20.8   & 19.2   & 16.3   & 11.9   & 9.7   & 7.0   & 5.2   \\[1mm] \cline{2-15} \rule{0mm}{4mm}
&\bar{\nc}& 85.6 & 130.9  & 257.8 & 291.8 & 239.5 & 239.3 & 140.8 & 94.6 & 61.9 & 35.1 & 22.5 & 13.0 & 11.0 \\[1mm] \hline\hline
\raisebox{-3mm}[0mm][0mm]{$\pi$} \rule{0mm}{4mm}&\mbox{fitness}& 32.4   & 41.5   &53.3 &52.7 & 39.6   &42.2 & 29.7   & 24.8   & 18.8   & 15.8   & 12.8   & 10.0   & 3.0   \\[1mm] \cline{2-15} \rule{0mm}{4mm}
&\bar{\nc} &  237.0  &  338.4 &  487.8 &  424.9 &  337.4 &  273.9 &  171.7 &  104.9 &   66.1 &   40.7 &   26.2 & 25.4 &   19.8\\[1mm] \hline
\end{array}
$$
It is interesting to read this table together with figure~\ref{codingst}. This confirm that TM with higher mutation rate 
need to limit their number of coding states. Moreover, the table and the figure make evident that equal fitness machines can largely  
differ in the numbers of coding states (even by a factor larger than $2$). 

So far we have examined the relation among the number of coding triplets and the fitness. Now we want to elucidate the relation among 
number of coding triplets $\nc $ and the number of total triplets $\nt=2 N$ (where $N$ is the total number of states). 
In the following table we report the ratio $\frac{\nc}{\nt}$. The values of $\nc$ and $\nt$ have been obtained 
averaging on the four values of mutation rate corresponding to the four best fitness scores for any value of $\pii$.
We excluded the other values of $\pmm$ from the averaging to avoid the already discussed limiting effect on the number of coding triplets 
that occur at extreme values of the mutation rate. 
\begin{table}[!h]
\caption{\label{tab-ratio}For the two tasks, we report the observed ratio of coding triplets versus the total number of triplets.}

\vspace*{-4mm}
$$\begin{array}{|c||c|c|c|c|c|c|c|c|c|c|c|c|}  \hline  \rule{0mm}{4mm} \mbox{task}
 & \pii & \frac1{ 10800} & \frac1{  6000} & \frac1{  3333} & \frac1{  1852} & \frac1{  1029} & \frac1{   572} & \frac1{   318} & \frac1{   176} & \frac1{    98} & \frac1{    54} & \frac1{    30}  \\[1mm]    \hline\hline \rule{0mm}{4mm}
\mbox{primes} &\frac{\nc}{\nt}&    0.262 &  0.255 &  0.183 &  0.139 &  0.091 &  0.098 &  0.041 &  0.044 &  0.030 &  0.021 &  0.019\\[1mm]
\hline\hline \rule{0mm}{4mm}
\pi & \frac{\nc}{\nt}& 0.403 & 0.243 &  0.258 &  0.231 &  0.154 &  0.077 &  0.081 &  0.066 &  0.054 &  0.032 &  0.026  \\[1mm] \hline
\end{array}
$$
\end{table}
As it is evident from the table, on the contrary of what happens for the fitness, the ratio $\frac{\nc}{\nt}$ basically decreases 
with the states-increase rate.

\begin{figure}[!h]
%\setlength{\unitlength}{1mm}
%\begin{picture}(150,60)
%\put(1,58){\includegraphics[angle=270,scale=0.6]{primes_nesoni_p2_loglog.ps}}
%\put(88,58){\includegraphics[angle=270,scale=0.6]{pi_nesoni_p2_loglog.ps}}
%\multiput(1,17)(86,0){2}{\begin{rotate}{90}number of coding triplets \end{rotate}}
%\put(38,0){(a)}\put(125,0){(b)}\put(65,5){$\pii$}\put(152,5){$\pii$}
%\end{picture}
\includegraphics{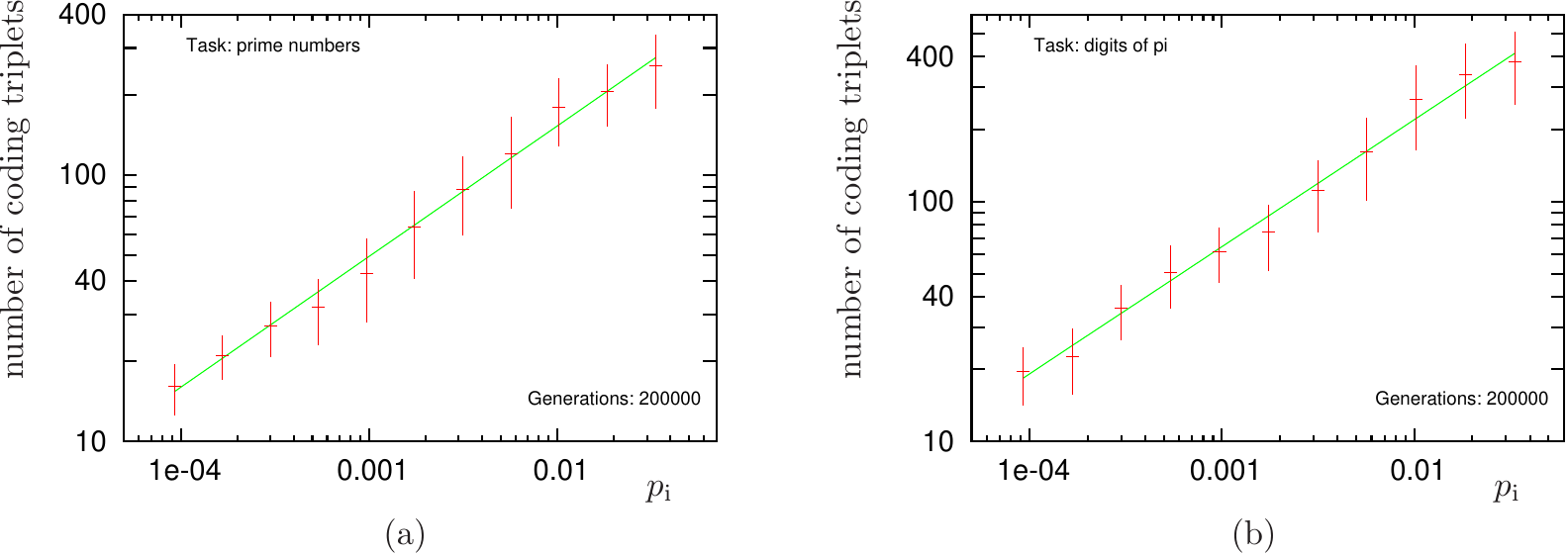}
\caption{\label{fit} In a log-log scale, we present the relation between the average number of coding triplets 
$\bar\nc$ and the states-increase rate at the last generation. We average $\nc$ on the four values of 
the mutation rate corresponding to the four best fitness values for each $\pii$.}
\end{figure}

By plotting the final number of coding triplets versus the states-increase rate on a log-log scale (in Figure~\ref{fit}), a 
power-law relation clearly emerges. By fitting the data, we obtained the following relation
\begin{eqnarray}
\nc &=& 1.5\cdot 10^3\ \pii^{0.49}  \quad \mbox{for prime numbers}  \label{nc1}\\
\nc &=& 2.5\cdot 10^3\ \pii^{0.53}  \quad \mbox{for digits of }\pi  \label{nc2}
\end{eqnarray}
On the other hand, the number of states has roughly a linear growth (see eq. (\ref{nstates})). This explains the 
decreasing behaviour of $\frac{\nc}{\nt}$ observed in the previous table.  

The last datum we want to present is the mean fitness increase $\bar{f}_{\text{inc}}$ namely the final fitness divided by the number of
increments. After averaging on the mutation rate, the states-increase rate and the seeds, we finally obtain
\begin{equation}
\begin{array}{cc}
\bar{f}_{\text{inc}} = 1.017 & \quad \mbox{for prime numbers}\\
\bar{f}_{\text{inc}} = 1.073 & \quad \mbox{for digits of }\pi 
\end{array} \label{mfitinc}
\end{equation}
Notice that the minimum theoretical value is 1, meaning that fitness always increases by one. Our values imply that
jumps in the fitness larger than one occur but are extremely rare. On this basis, we would say that the evolution of
our TMs is more gradualist than saltationist. This is, in some way, surprising since mutations occurring in the state-call entry
of the coding triplets almost always give raise to a radical mutation of the TM.

\subsection{\label{comparative}A comparative run}

There are two basic arguments that could lead the reader to think that the fitness has to be always a monotonic 
increasing function of $\pii$ (for all values of $\pii$, regardless of the choice of the target tape). 

The first argument is that the fitness growth with $\pii$ has to be considered expected and obvious \cite{luke} because 
the higher the number of states is, the higher is the number of TMs that solve the given task (goal tape). 
Moreover, each TM can be thought to be contained in a bigger one just by adding some non-coding states 
while the opposite is clearly false. 

The second argument is that, since there is no direct cost associated to the accumulation of non-coding triplets,
while, as we will argue in the next section, there is an indirect advantage, again the fitness should always be a monotonic increasing function of $\pii$. 

The considerations in the first argument are absolutely correct, but they apply also to bad genomes:  there are more large bad genomes than small bad genomes. Given that the system can only experiment a limited (and fixed) number of trials, having a larger set of TMs within which to search just implies that a larger number of trials will fail.
In other words, increasing the number of states means that there are more good solutions but does not mean
that it is easier to find them\footnote{On the contrary, it is natural to think that from some number of states on, it will be extremely more difficult to find them.}.

Our answer to the second argument is that there is also an indirect cost associated with a too 
fast accumulation of non-coding triplets. This cost is strictly related to the argument given before.
Indeed, suppose that there is an optimal solution represented by a TM with $M$ coding triplets. 
If, at a given generation, the maximum number of states in the TMs population is $N$, during the successive generations
the evolutive algorithm will test TMs with a number of coding states distributed between $1$ and $N$. If $N<M$, there 
is no chance of finding the optimal solution. On the other hand from some value of $N>M$ on, the number of TMs 
with $M$ coding triplets that will be tested, at each generation, will progressively decrease. In other words,  
the value of $\pii$ determines the ``time'' that the system has to explore the set of TMs with a given number of coding triplets. 
If $\pii$ is very large, the system will try a lot of TMs with many coding triplets, risking to miss some
optimal solution with fewer ones. We think that this should be a general phenomenon in our evolutive model 
for any choice of the goal tape. In other words, we think that there is both an indirect advantage and an indirect cost associated with the accumulation of non-coding triplets,
whose relative weights depend on $\pii$.
Since the maximum fitness is bounded, the indirect advantage must decrease from some value of $\pii$ on; 
viceversa, the indirect cost will reasonably increase. In conclusion we expect that, in our evolutionary model
for TMs, the fitness will have a maximum for some particular value of $\pii$, depending on the particular choice of 
the goal tape. Here we present the result of some simulations where the goal tape and the number
of TMs in the population have been chosen in such a way that the maximum of the fitness is obtained for a value
of $\pii$ that we can computationally afford.

In particular we choose a goal tape with ones in the multiples of $5$ positions. There is an obvious $5$-states 
TM that solve this task:

\begin{center}
\begin{tabular}{|c|c|c|c|c|c|}
\hline  &\bf 1 &\bf 2 &\bf 3 &\bf 4 &\bf 5  \\
\hline 0 &  0-Right-\bf 2 &  0-Right-\bf 3 & 0-Right-\bf 4 &  0-Right-\bf 5 & 1-Right-\bf 1 \\
\hline 1 & \_--\_--\_ & \_--\_--\_ & \_--\_--\_ & \_--\_--\_  & \_--\_--\_  \\ \hline
\end{tabular}
\end{center}

\noindent We also took a very small population of 20 individuals, we limited calculations to 1000 generations and machines were stopped at 400 time steps. The results of our simulations are summarized in table~\ref{multipli}. 

\begin{table}
\caption{\label{multipli}
We report the average fitness obtained after 1000 generations, with a population of 20 individuals,
$20$ different choices of the seeds and multiples of 5 as goal tape.} 
\begin{center}
\begin{tabular}{c|c|c|c|c|c|c|c|c} 
\cline{2-8} &  & 0.0025    & 0.005    & 0.01    & 0.02    & 0.04    & 0.06 &$\pmm$   \\ \cline{2-8}
& 0.0022& 0.1   & 0.1   & 0.3   & 0.2   & 0.5   & 0.1   \\  \cline{2-8}
& 0.005& 0.1   & 0.2   & 2.3   & 0.6   & 0.5   & 0.1   \\  \cline{2-8}
& 0.022& 0.0   & 3.1   & 3.2   & 2.8   & 0.7   & 0.1   \\  \cline{2-8}
& 0.05& 0.0   & 3.2   & 14.0   & 7.4   & 0.3   & 0.0   \\  \cline{2-8}
& 0.1& 0.0   & 0.1   & 10.2   & 2.9   & 2.1   & 0.0   \\  \cline{2-8}
& 0.22& 0.1   & 0.0   & 5.5   & 6.4   & 0.2   & 0.0   \\  \cline{2-8}
& 0.5& 0.0   & 2.1   & 3.1   & 0.2   & 0.1   & 0.0   \\  \cline{2-8}
$\pii$& 1& 0.0   & 4.5   & 3.5   & 0.1   & 0.0   & 0.0   \\  \cline{2-8}
\end{tabular}\end{center}
\end{table}

As is evident from table~\ref{multipli}, the fitness initially grows for small values of $\pii$, then saturates and then decreases at higher values. Similar results have been obtained using as goal tapes the multiples of 2 and 4 
(data not shown).  

We observe that, for the multiples of 5 goal tape, the TMs use a  small number of coding triples (on average, from 5 to 10 triplets), quite unrelated to the fitness or to the mutation and states-increase rates.
The reason is that the TMs tend to imitate the 5-states machine indicated earlier (or an equivalent one).
On the opposite, in the other simulations we observed a clear relashionship between fitness and number of coding triplets (see fig.~\ref{fit}) that indicates that, in general, the TMs that we obtain do not encode 
periodic (or other compact) algorithms, but try to ``guess'' a $1$ after another (as follows also by the fact
that the mean fitness increase is very near to $1$, see eq. \ref{mfitinc}).

%%%%%%%%%%%%%%%%%%%%%%%%%%%%%%%%%%%%%%%%%%%%%%%%%%%%%%%%%%%%%%%%%%%%%%%%%%%%%%%%%%%%%
\section{Conclusions and discussion}

We developed an abstract model, mimicking biological evolution, to understand if there is an ``evolutive'' advantage in
maintaining non-coding parts in an algorithm. 
We tried to keep the model the simplest as possible, while being complicated enough not to allow easy predictions. 
Moreover we were requiring a model where it would be easy to distinguish between coding and non-coding states in an 
unambiguous way, with a simple mechanism for
accumulation of non-coding states, where mutation rate and states-increase rate should be independent,
with a simple mechanism of state activation through mutation. The use of TMs fitted perfectly with these requirements. 

For the sake of simplicity, we imposed various restrictions on our model that can be relinquished to make the model 
more realistic from a biological point of view. In particular we decided that:
\begin{enumerate}
\item non-coding states accumulate at a constant rate (determined by the states-increase rate
$\pii$) without any deletion mechanism,
\item there is no selective disadvantage associated with the accumulation of both coding and non-coding states, 
\item the only mutation mechanism is given by point mutation and it also occurs at a constant rate (determined by the mutation rate $\pmm$),
\item there is a unique ecological niche (defined by the target tape), 
\item population is constant,
\item reproduction is asexual.
\end{enumerate}     
Because of the second point, the TMs we have obtained are not economical in the use of coding triplets. Indeed it is
very easy to think of TMs with many fewer coding triplets reaching a better fitness. 
It could be interesting to make further studies trying to relinquish some of these restrictions.
However, letting fall any of these restrictions, would have introduced further free parameters in our model and this
was undesirable for two reasons. Firstly the number of needed simulations increases with the number of free parameters
and secondly it becomes more difficult to interpret unambiguously the results. 

We decided to use asexual reproduction since
crossover would have introduced other possible causes for the accumulation of non-coding states 
(as in the ``hitchhiking'' and ``defence against crossover'' explanations of code-bloat) blurring the final conclusions. 
Analogously, the processes of insertion and deletion would have
brought the ``removal bias'' effect. None of this processes has any role in our model. Needless to say, also the  
processes advocated by the biological explanations (``junk DNA'', ``selfish DNA'',
``nucleoskelethal'' and ``nucleotypical'' theory) that we briefly sketched in the introduction, are ineffective
in our model.

In our simulations we started from an initial population completely unadapted to its ecological niche and we observed
that, for fixed $\pmm$, higher values of $\pii$ correspond to a faster evolution. This behaviour is observed for both 
tasks on the whole range of generations (see figures \ref{storia_primes}, \ref{storia_pi}). The same behaviour 
is observed also for the number of coding triplets; the striking similarity between figures \ref{plotprimi}.a, 
\ref{codingst}.a and \ref{plotpi}.a, \ref{codingst}.b strongly suggests a direct link between the number of coding
triplets and the fitness reached at the end of the evolutive runs. Since the number of coding states is a monotonic
increasing function of $\pii$ (see figure \ref{fit}), the natural conclusion is that TMs evolving under higher values
of the state-increase rate $\pii$ reach a better fitness since they succeed in generating a larger number of coding 
triplets. From table \ref{tab-ratio} we see that even if the number of coding triplets grows, they progressively 
become a very small fraction of the genome as $\pii$ increases
(this fact also easily follows from the observation that the number of coding triplets grows approximately as the 
square root of $\pii$ (see eq. \ref{nc1}, \ref{nc2}), while the total number of states grows as a linear function 
of $\pii$ (see eq. \ref{nstates})). So, we argue that the dependence of the fitness on the state-increase rate $\pii$
is realized through the following chain of implications: larger values of $\pii$ implies a larger availability 
of non-coding triplets that, in its turn, implies a greater probability of enlarging the number of coding triplets. 
We give the following explanation of the latter statement. During evolution, a mutation occurring
in the state-call entry of a coding state can activate a certain subset of non-coding states. Since this activation
is probably nefarious, such mutations will almost always lead to the extinction of the TM. Having a population of
TMs with a large number of non-coding states allows the system to try to activate a large number of different 
subset of them (by means of different mutations in the state-call entries) until a good subset is found. 
Such explanation is supported by the fact that the final number of states is a selected character 
(at least for low values of $\pii$) as discussed in section \ref{results}. 

As we discussed in section~\ref{comparative}, we expect that the fitness should have a maximum around some
value of $\pii$ and it would be very interesting to determine its actual position for the primes
and $\pi$ goal tapes. 
In such cases, the maximum must lie outside the range of values of $\pii$ that we have considered. 
Moreover, it is worthwhile to notice that from figures \ref{plotprimi}.c, \ref{plotpi}.c, we seem to be still quite
far from this maximum. This means that the optimal coding/non-coding ratio, in our model, is probably much less than
$2 \%$. Exploring the region $\pii>1/30$ could provide interesting 
informations but is computationally quite expensive.

\subsection{Back to biology} 
In this section we put forward some biological speculations inspired by our model.   
There are two way of identifying TMs with biological entities and they suggest two ways up to which the accumulation 
of non-coding free to mutate DNA can play a role for ``evolvability''.   
In the first one we identify TMs with organisms and coding-states with
genes. We have to stress that the mechanism of transcription is different in the two contexts. 
For TMs transcription is serial, so that states must be transcribed, one at a time, in a prescribed order, while
in biological organisms transcription of genes can happen in parallel. We can interpret TMs states as genes accomplishing
both a structural and regulatory function, since a coding state both affects the output tape and specifies which state
has to be successively transcribed.
From this point of view, we can think of TMs in our simulations as organisms trying to increase their
gene pools adding new genes assembled from junk DNA. If the organisms possess more junk DNA it is possible
to test more ``potential genes'' until a good one is found. 

On the other hand we can identify the TMs with single genes and their states as sequences of nucleotides. 
From this point of view, transcription of states is serial as the transcription of the nucleotides composing
a gene in DNA. The difference now is that transcription
of states can jump forward and backward (respect to the natural order of the states) and pass also more than one
time through a given state), while transcription of a gene proceeds always in the 5'-3' direction. 
Letting these differences apart, from this point of view we think of our simulations as the assembly of a new gene
by the addition of new nucleotide sequences. In our explanation of why TMs with a large number of non-coding states
evolve faster, the possibility of jumping in the transcription is essential. Indeed, without this possibility,
the TMs could test only the subset of non-coding states just following the last coding state.
We notice that this is exactly what happens in prokaryotes, while in eukaryotes the splicing of introns allows 
to have jumps in the transcription (at least in the forward direction). So, in the framework of
this second interpretation, our model suggests
that the mechanism of splicing could have a very significant role for the evolvability of eukaryotes.

\section*{Acknowledgements}
It is a pleasure to thank M. Caselle, A. Parmeggiani, G. Satta and S. Pagnotta
for their helpful comments regarding this work and for having read the manuscript.
GF thanks INFN for a post-doctoral fellowship and for financial support.
F.M. acknowledges a three months post-doc fellowship at LAPTH by the
research training network "EUCLID. Integrable models
and applications: from strings to condensed matter", contract number
HPRN-CT-2002-00325.
We thank the "Centre de calcul IN2P3-CNRS" and the ``EUMEDGRID'' INFN project for having allowed us run our simulations
on their grid facilities.

%%%%%%%%%%%%%%%%%%%%%%%%%%%%%%%%%%%%%%%%%%%%%%%%%%%%%%%%%%%%%%%%%%%%%%%%%%%%%%%%%%%%%%%%%%%%%%%%%%%%%%%%%%%%%%%%%%%%%%

\end{document}